\documentclass[
reprint,
superscriptaddress,
 amsmath,amssymb,
aip,
pop,
floatfix,
]{revtex4-1}

\usepackage{graphicx}
\usepackage{epstopdf}
\usepackage{dcolumn}
\usepackage{bm}
\usepackage{ifpdf}
\usepackage[usenames,dvipsnames]{color}
\usepackage{hyperref}
\def\Snospace~{\S{}}

\usepackage[all]{hypcap}
\hypersetup{urlcolor=Blue, citecolor=Blue,linkcolor=Blue,colorlinks=true}

\newcommand{\sref}[2]{\hyperref[#1]{Fig. \ref{#1}#2}}

\newcommand{\Fig}[1]{
	\label{fig:#1}
	\includegraphics[scale=1]{plasmagun2018_figures_#1-eps-converted-to.pdf} 
}

\newcommand{\x}[1]{$x=#1$ mm}

\newcommand{\z}[1]{$z=#1$ mm}

\newcommand{\xcmcubed}[2]{$#1\times10^{#2}$ cm$^{-3}$}
\renewcommand{\deg}[1]{#1$^{\circ}$}

\begin{document}

\title{Two-Colour Interferometry and Thomson Scattering Measurements of a Plasma Gun}

\author{J. D. Hare}
\email{jdhare@imperial.ac.uk}
\affiliation{Blackett Laboratory, Imperial College, London, SW7 2AZ, United Kingdom}
\author{J. MacDonald}
\affiliation{Blackett Laboratory, Imperial College, London, SW7 2AZ, United Kingdom}
\affiliation{AWE Aldermaston, Reading, Berkshire, RG7 4PR United Kingdom}
\author{S. N. Bland}
\author{J. Dranczewski}
\author{J. W. D. Halliday}
\author{S. V. Lebedev}
\author{L. G. Suttle}
\author{E. R. Tubman}
\affiliation{Blackett Laboratory, Imperial College, London, SW7 2AZ, United Kingdom}
\author{W. Rozmus}
\affiliation{High Energy Density Science Division, SLAC National Accelerator Laboratory, Menlo Park, California 94025, USA}
\affiliation{Theoretical Physics Institute, Department of Physics, University of Alberta, Edmonton, Alberta, Canada T6G 2E1}

\date{\today}

\begin{abstract}
	
We present experimental measurements of a pulsed plasma gun, using two-colour imaging laser interferometry and spatially resolved Thomson scattering. Interferometry measurements give an electron density $n_e\approx2.7\times10^{17}$ cm$^{-3}$ at the centre of the plasma plume, at 5 mm from the plasma gun nozzle. The Thomson scattered light is collected from two probing angles allowed us to simultaneously measure the collective and non-collective spectrum of the electron feature from the same spatial locations. The inferred electron densities from the location of the electron plasma waves is in agreement with interferometry. The electron temperatures inferred from the two spectra are not consistent, with $T_e\approx 10$ eV for non-collective scattering and $T_e\approx 30$ eV for collective scattering. We discuss various broadening mechanisms such as finite aperture effects, density gradients within the collective volume and collisional broadening to account for some of this discrepancy. We also note the significant red/blue asymmetry of the electron plasma waves in the collective scattering spectra, which could relate to kinetic effects distorting the distribution function of the electrons.

\end{abstract}

\maketitle
\section{Introduction}

Laser based diagnostics provide detailed information on plasma parameters through measurements of the phase or spectrum of light passing through the plasma.
Using diagnostics which work on fundamentally different physical principles provides a more complete picture of a given plasma, and allows results and analysis to be cross checked, validated and constrained.
In this paper we present and compare results from laser interferometry and Thomson scattering used to study a relatively simple plasma from a pulsed source.

In laser interferometry,  the phase difference between a probing beam, which passes through a plasma, and a reference beam, which passes around a plasma, can be used to infer the refractive index of the plasma.
In cold plasmas with low ionisation fractions, this refractive index can include significant contributions from both the neutral refractive index (which increases with wavelength) and the electron refractive index (which decreases with wavelength).
Two-colour interferometry is a technique which exploits these different dependencies of refractive index on probing wavelength by combining phase maps from two interferograms produced at the same time at two distinct wavelengths.\cite{Muraoka2000}
This technique is therefore used to separate the electron and neutral contribution and give separate estimates for the electron and neutral densities, and hence can provide additional constraints when analysing other diagnostics such as Thomson scattering.

Thomson scattering is a powerful tool for studying the properties of plasmas over a vast range of densities and temperatures, from cold, low temperature industrial plasmas\cite{Carbone2015} to magnetic confinement devices\cite{Peacock1969} and high energy density plasmas.\cite{Swadling2014a}
Unlike many line integrated diagnostics, such as self-emission imaging, spectroscopy, interferometry and Faraday rotation imaging, Thomson scattering provides a truly local measurement of the plasma properties inside a small volume defined by the focal spot and the collection optics.

Thomson scattering operates in two principle regimes --- collective scattering, in which the wavelength of the scattering light fluctuations is larger than the Debye length ($\lambda_{De}=\sqrt{\epsilon_0 k_B T_e/n_e e^2}$) and the light scatters from collective modes within the plasma, and non-collective scattering in which the wavelength is smaller than $\lambda_{De}$ and hence the laser scatters from individual electrons, whose motion is uncorrelated.

In the collective regime, the scattered light at a given angle can be analysed to determine the energy and momentum of these resonant collective modes, which provides information on the plasma conditions, such as the electron and ion temperatures, the electron density and the bulk velocity in the direction of the resultant scattering vector.
In the non-collective regime, the scattering spectrum is a direct measurement of the distribution function of the electrons or ions, though the noise in a realistic system limits measurements to velocities close to the thermal velocity.

Both the non-collective and collective scattering regimes can sometimes be obtained in the same experiment when the scattered light is collected from two different directions.
If the light is collected from the same plasma volumes, the parameters inferred from both regimes should be consistent.
However, in any real experimental set-up there are additional effects on the spectrum which must be carefully accounted for in order to reach agreement.
These effects may be significant in experiments which collect the scattered light from only one direction, and therefore have no inherent checks on the validity of the plasma parameters inferred from Thomson scattering.

In addition to cross checking results from Thomson scattering spectra taken at different angles, it is important to place additional constraints during fitting the spectra, as the models used are highly non-linear.\cite{Kasim2018a}
These constraints can take the form of reasonable heuristics about possible densities and temperatures, or information taken from other diagnostics.
In this paper we will demonstrate how we used two-colour interferometry to produce maps of the electron density, which can be compared with Thomson scattering fits obtained simultaneously at two different scattering angles.
Although we observe good agreement between the electron densities inferred from the interferometry and Thomson scattering, the temperatures inferred from the two Thomson scattering spectra disagree considerably, and we consider some additional broadening mechanisms which partially account for this discrepancy.

This paper is organised as follows: in \autoref{sec:experimental_setup} we describe the set up of the plasma gun and the diagnostics: laser interferometry and Thomson scattering. 
In \autoref{sec:results} we present results from two-colour interferometry, which gives unphysical negative neutral atom densities in some regions of the plasma jet, but fortunately not in the region in which we perform Thomson scattering. 
We then present Thomson scattering spectra, and show that we can resolve electron plasma waves in the collective and borderline non-collective regime. 
In \autoref{sec:discussion} we compare the electron densities obtained from interferometry and Thomson scattering, and show there is good agreement. 
We also present the temperatures measured by Thomson scattering at two different collection directions, which give wildly different results. 
We propose some possible broadening mechanisms, including finite aperture effects, density gradients within the collection volume, and collisional broadening, which together account for a significant proportion of the inferred electron temperature, though not the full discrepancy measured.
Finally we look at the blue/red asymmetry in the intensity of the electron plasma waves, and discuss the possible role of heat transport and non-kinetic effects in this plasma.

\section{Experimental Setup}\label{sec:experimental_setup}

\begin{figure}[t]
	\Fig{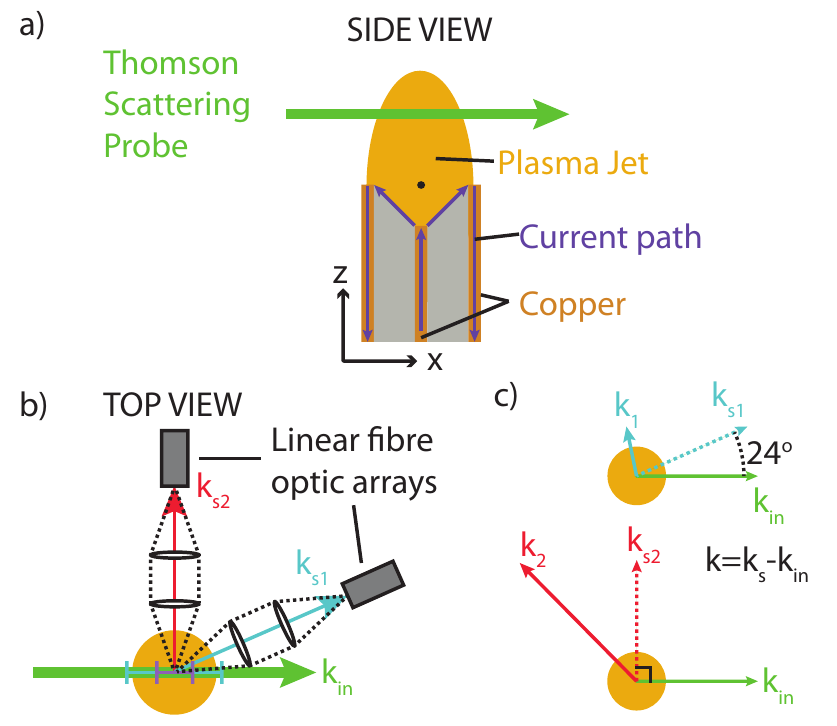}
	\centering
	\caption{Experimental set-up. a) Side view of the plasma gun, showing a cross-section of the co-axial cable and the current path. Ablation of the dielectric and copper produces a plasma jet. The Thomson scattering probe passed along a horizontal chord through the centre of the plasma jet, at \z{5}. The origin of the coordinate system is marked with a black dot. b) Top view of the plasma gun, with the jet coming out of the page. The Thomson scattered light is observed using two fibre optics arrays using a two-lens telescope. The positions of the fibre arrays defines two scattering vectors, $\textbf{k}_{s1}$ and $\textbf{k}_{s2}$. c) The resultant scattering vectors $\textbf{k}=\textbf{k}_{s}-\textbf{k}_{in}$ for fibre arrays at \deg{90} and \deg{24} to the probe beam.}
\end{figure}

\subsection{The Plasma Gun}

The plasma gun used in these experiments was originally designed for use in plasma opening switches, providing a simple, low jitter plasma source. 
The gun is made from copper semi-rigid coaxial cable, with an outer diameter of 6.25 mm and a Teflon dielectric insulator (\sref{fig:setup}{a}).
The end of the cable is machined to produce an inverted cone with an angle of \deg{60},\cite{Goyer1993} and is driven using a 0.7 $\mu$F capacitor charged to 21 kV.
As the capacitor discharges, the dielectric insulator breaks down, allowing current to flow across the dielectric surface.
The current pulse rings, forming a decaying sinusoid with a peak current of 20 kA and a period of around 4 $\mu$s  --- the data presented in this paper was taken at 940 ns after current start, near the maximum current.
The surface plasma is formed by the ablation of material from the Teflon insulator and the copper conductors by the current pulse, and is therefore mostly carbon and hydrogen with some copper and fluorine impurities.

This plasma gun has previously been studied using multi-time, single colour interferometry and time-resolved single chord two-colour triature interferometry.\cite{Macdonald2015}
These diagnostics show that the plasma accelerates from the surface, reaching velocities of around 1 km/s as the plume expands into vacuum.
The electron density rapidly drops away from the nozzle of the plasma gun, initially at $n_e=$ \xcmcubed{1}{19} but reaching \xcmcubed{1}{17} after a few mm.
Spectroscopy suggests that the plasma is known to be relatively cold ($T_e=$ 5 to 10 eV) with a significant neutral gas density.

\subsection{Diagnostic Set-up}

In this paper, we extend this previous work using two diagnostics: a two-colour spatially and temporally resolved Mach-Zehnder interferometer, and spatially and temporally resolved Thomson scattering of the collective and non-collective electron feature.

We used Mach-Zehnder laser imaging interferometry to measure the line integrated electron and neutral density in the side-on ($z-x$) plane. 
Interferometry was carried out using the 2nd (532 nm) and 3rd (355 nm) harmonics of a Nd-YAG laser (EKSPLA SL321P, 500 ps, 500 mJ)\cite{Swadling2014a}, with the beams imaged onto Canon 500D DSLR cameras.
The 3rd harmonic was delayed by 20 ns with respect to the 2nd harmonic, but this time-scale is short compared to the experimental time-scales and so it is reasonable to treat the two-colour probing as simultaneous.

The shutters of the DSLR cameras were left open for 1.3 seconds, allowing them to be triggered before the application of the current pulse.
The laser energy was set such that the laser intensity significantly exceeded the intensity of the self-emission from the plasma --- long exposure self-emission images (not shown in this paper) suggest that the time-integrated self-emission was three orders of magnitude smaller than the laser light.
The probing paths for the two harmonics were identical, with the beams combined using a dichroic beamsplitter, travelling through the same optics until they were split with a second dichroic beamsplitter just before the DSLR cameras.

The fringe shift in the interferograms is due to contributions to the phase from the free electrons, the neutrals and the ions.
These contributions are line integrated, and due to the azimuthal symmetry of the plasma gun we are able to use Abel inversion to estimate the phase shift as a function of radius. 
The phase shift due to free electron density is linearly proportional to the laser probing wavelength, and the phase shift due to neutrals is inversely proportional to the wavelength. 
By solving simultaneous equations for the phase shifts at the two different wavelengths, we can obtains maps of the electron and neutral densities.
As we will show later in this paper, this simple treatment does not always give physical results, though the reasons for these discrepancies are unclear.

Thomson scattering was used to measure the electron temperature and electron density in spatially resolved locations across the plasma jet.
A laser beam (532 nm, 2 J, 8 ns) was focused using a 750 mm achromat through the plasma, and the scattered light was imaged onto two fibre optics arrays at different scattering angles (\sref{fig:setup}{b}) by two two-lens telescopes which sit inside the vacuum chamber.
Each telescope consists of two achromat lenses (50 mm diameter, 150 mm focal length), with the first at 150 mm from the plasma, and the second lens a further 150 mm away and connected by lens tube to avoid any stray light entering the optical set up.
The second lens focuses light onto a linear fibre optic array located outside of the vacuum chamber. 
The two arrays consist of seven evenly spaced 200 $\mu$m diameter fibres, which are separated by an identical ``dead'' fibre, leading to a 400$\mu$m centre-to-centre spacing which reduces ``cross-talk'' between fibres when imaged onto a CCD.
These arrays therefore collect light from seven spatially distinct volumes within the plasma.

The light from the fibres is combined onto a single 14 fibre linear array, which is imaged by a two-lens telescope onto the entrance slit of an Andor Shamrock 500i Czerny-Turner spectrometer using a grating with 1200 lines/mm.
The entrance slit width was chosen as a compromise between the signal to noise ratio and spectral resolution, with a slit width of 120 $\mu$m for the results presented in this paper, which corresponds to a Voigt-profile response function with FWHM of 2\AA{}.
The light is collected by an intensified CCD (Andor iStar 334) with a 4 ns gate, triggered by a signal from a diode which sees a fraction of the Thomson beam earlier in the amplifier chain.

Although the ion feature dominates the Thomson scattering spectra, for this plasma gun the bulk and thermal motions provide only small Doppler shifts and Doppler broadening, which cannot be resolved with this spectrometer.
Instead, we chose to saturate the ion feature on our ICCD and study the electron feature of the Thomson scattering spectra.
The electron modes are high frequency, which corresponds to a larger Doppler shift, and so can easily be distinguished from stray light at the initial laser wavelength.

Scattering can occur in the collective or non-collective regimes, determined by $\alpha=1/|\textbf{k}| \lambda_{De}$ where $\textbf{k}=\textbf{k}_s-\textbf{k}_{in}$ is the wave--vector of the mode which couples $\textbf{k}_{in}$ to $\textbf{k}_s$ and $\lambda_{De}$ is the electron Debye length. 
As Thomson scattering is an almost elastic process, $|\textbf{k}_s|\approx|\textbf{k}_{in}|= k_{in}$, and so $|\textbf{k}|=k\approx2 k_{in} \sin{(\theta/2)}$, where $\theta$ is the angle between the incoming laser beam and the scattered light.
For a plasma at a specific temperature and electron density, probed by a set wavelength, there is only one free parameter left to control the value of $\alpha$ and hence determine whether the scattering is collective or non-collective, and this is the angle from which the scattered light is collected.

In the collective regime, the wavelength of the scattering fluctuations is larger than the Debye length, and hence the free electrons react collectively.
The high frequency mode is the Electron Plasma Wave (EPW) or warm longitudinal plasma wave, which follows the Bohm-Gross dispersion relationship:
\begin{eqnarray}
\omega^2&=&\omega^2_{pe}+3 k^2 v^2_{th,e}\\
&=&\omega^2_{pe}\left(1+3/\alpha^2\right) \label{eqn:eqw_dispersion}
\end{eqnarray}
where $\omega^2_{pe}=e^2 n_e/\epsilon_0 m_e$ is the electron plasma frequency.
Hence the position of the resonances, at a frequency shift of $\pm\omega$, depends on both the electron density and the electron temperature.
For a given spectrum in the collective regime, it is difficult to uniquely specify both the electron density and the electron temperature within experimental uncertainty --- this means is it is useful to have measurements from at least two angles, which allow the measurements to be constrained.
In our experiment, the spectra at \deg{24} are at large $\alpha\approx4$, where the position of the EPW resonance most strongly depends on the electron density alone, and the spectra at \deg{90} have an intermediate $\alpha\approx1$, where the EPW resonance is more sensitive to the electron temperature. 
In this paper we will refer to these two fibre arrays as $k_{s1}$ and $k_{s2}$, which have scattering fluctuation vectors $k_1$ and $k_2$ respectively.

\section{Results}\label{sec:results}

All the results presented in this paper are from a single shot.
The interferometry and Thomson probing lasers simultaneously collected data at 940 ns after the start of the current pulse.
We present results from this shot because the plasma conditions provided a measurable fringe shift on both interferometers, and the Thomson scattering spectra were in a regime which allowed us to infer both the electron density and electron temperature.
During this shot series, we also gathered interferometry and Thomson scattering data from 350 ns to 1710 ns after current start, and the analysis of these results will be the subject of future work.

\subsection{Interferometry: Electron and neutral densities}

\begin{figure*}[!t]
	\Fig{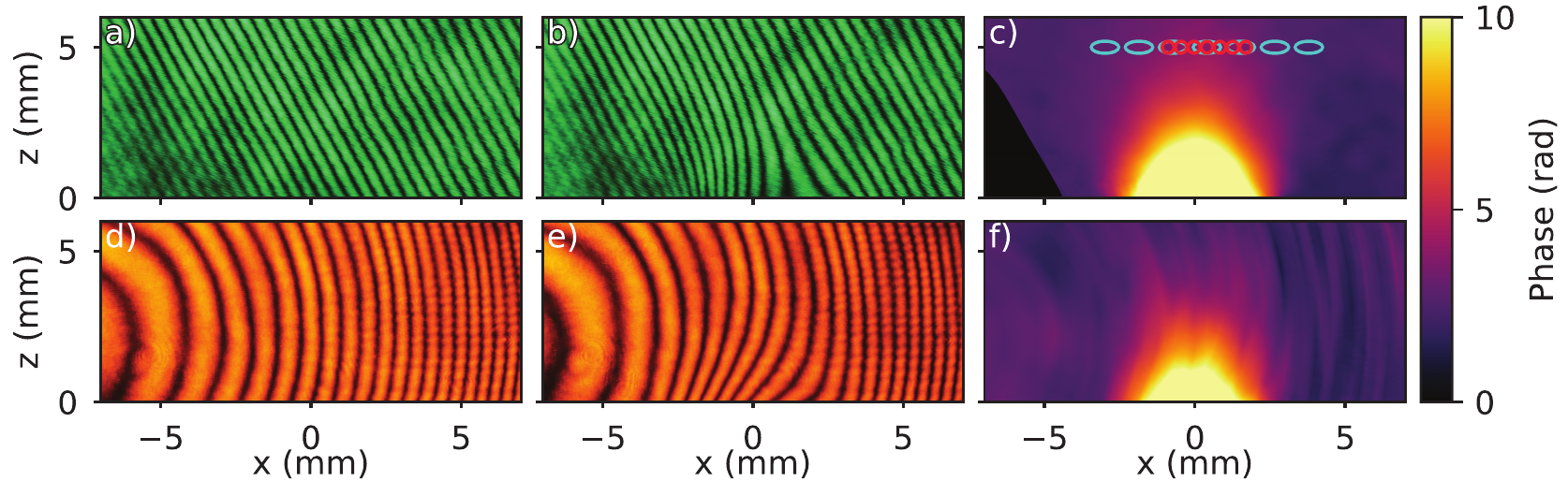}
	\centering
	\caption{Results from the two-colour interferometry diagnostic. The nozzle of the plasma gun is at \z{0} and between \x{\pm3}. 
		a) Vacuum interferogram at 532 nm taken in the absence of any plasma. 
		b) Interferogram from a 532 nm laser at 940 ns after current start. The fringes near the origin (bottom, centre) are clearly shifted compared to the fringes at the top or sides.
		c) The phase map computed by interpolation of the isophase contours in b) and subtracting the background phase. The locations of the Thomson scattering volumes are shown in cyan ($k_1$) and red ($k_2$), to scale.	d) Vacuum interferogram from a 355 nm laser in the absence of any plasma. 
		e) Interferogram from a 355 nm laser at 960 ns after current start. The fringes are also shifted near \x{0}. f)  Phase map computed from e), showing significant artefacts due to large, curved fringes.}
\end{figure*}

Results from the two-colour Mach-Zehnder imaging interferometry are presented in \sref{fig:interferometry_results}.
The raw interferograms (\sref{fig:interferometry_results}{a, b and d, e}) show the distortion of the interference fringes caused by the increased electron and neutral density caused by the plasma gun.
Away from the nozzle of the gun [at $(x,z)=(0,0)$], the fringe shift decreases, and near the corners of the image, the fringes are the same as in the interferograms taken in the absence of the plasma (the ``reference'' or ``vacuum'' interferograms seen in \sref{fig:interferometry_results}{a and d}.)
To determine the line-integrated phase along the probing laser direction, it is necessary to determine the phase shift between a vacuum interferogram and a shot interferogram, and we discuss this process in more detail in Appendix \ref{sec:magic}; further examples demonstrating the fidelity of this procedure can be found in Refs. \onlinecite{Swadling2014a} and \onlinecite{Swadling2013}.

The interpolated phase maps are shown in  \sref{fig:interferometry_results}{c and f}), produced by manually tracing the constructive and destructive interference fringes using Adobe Photoshop.
The phase map for the 532 nm interferometry (\sref{fig:interferometry_results}{c})  is quite smooth, and appears to be left-right symmetric, making it suitable for Abel inversion.
The data is cut off in the bottom left corner due to the reduced fringe contrast making it difficult to reliably reconstruct the phase map.
Thomson Scattering collection volumes for arrays $k_{s1}$ (cyan) and $k_{s2}$ (red) are overlaid on this phase map, along the chord \z{5}.
The different fibre spacings are due to different magnifications for the two arrays, caused by the projection of the fibre array onto the probing laser beam from different angles.
The collection volumes for $k_1$ are elongated along the beam and collect light from an elliptical volume.

The interpolated phase map for the 355 nm interferometry (\sref{fig:interferometry_results}{f}) contains significant artefacts due to the large fringe spacing, which makes it difficult to track the exact maxima or minima of the fringes. 
This leads to different interpolated phase maps between the reference and shot interferograms, and the artefacts occur during the subtraction step.
These artefacts make this phase map unsuitable for numerical Abel inversion, as the artefacts will strongly distort the final result.

In order to determine the electron and neutral density independently, we need to use the phase maps from both the 532nm and 355 nm interferometry.
This means that we need to find a way of modelling the data from the 355 nm interferometry that gives a smooth, left/right symmetric function.
We initially fit each row of pixels with a Gaussian profile multiplied by polynomials with only even powers\cite{Deutsch1983}, but found that only the lowest order polynomial (that is, a simple Gaussian profile fit) was significant, and so discarded the higher order terms.
The profiles are by construction smooth and symmetric, tending to 0 as $x\rightarrow\pm\infty$, as required for Abel inversion.

\begin{figure*}[t]
	\Fig{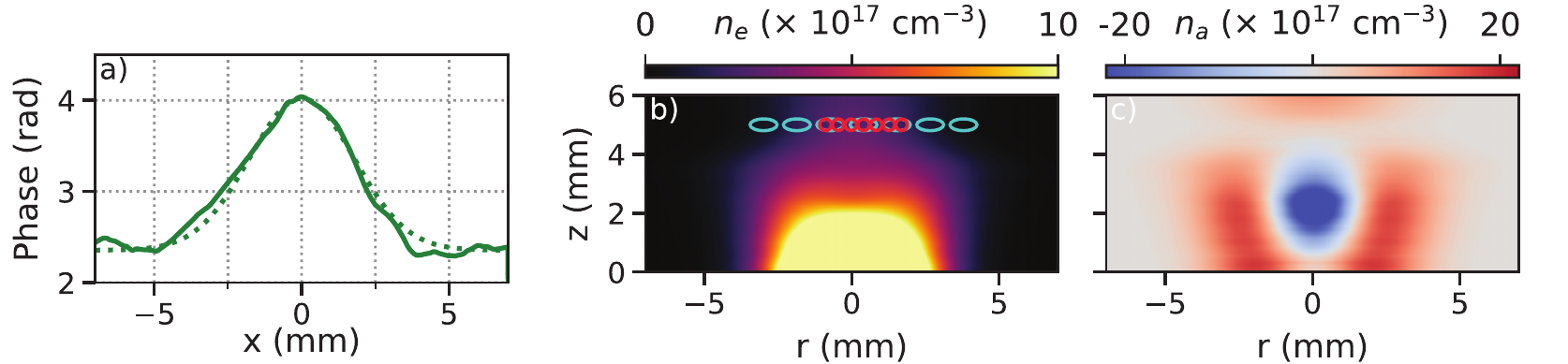}
	\centering
	\caption{Electron and neutral densities obtained via Abel inversion. a) Line-out of phase along \z{5} from \sref{fig:interferometry_results}{c} (solid) and fit with a Gaussian distribution (dashed). b) Electron density obtained by analytical Abel inversion of Gaussian fits followed by simultaneously solving for electron and neutral density using phase maps from 532 nm and 355 nm interferometry. c) Neutral density - note significant region of unphysical negative neutral density around \z{2}.}
\end{figure*}

We perform the Abel inversion using the analytical result for the transform of a Gaussian profile.
The suitability of the Gaussian profile fit is demonstrated in \sref{fig:electron_density}{a}, where the dashed green fit is a good approximation to the solid green line-out along \z{5} in \sref{fig:interferometry_results}{c}.
The smoothed phase maps are first analytically Abel inverted row-by-row using
\begin{eqnarray}
A\cdot G(x,\sigma)+ C\rightarrow (A/\sigma) \cdot G(r,\sigma)
\end{eqnarray}
where $G$ is the standard Gaussian or normal distribution, A is a scaling factor and C is an additive constant which accounts for any non-zero phase shift due to vibrations in the interferometry set-up, and which vanishes during the Abel inversion.
Note the change in the horizontal coordinate from $x\rightarrow r$, which is reflected in the relabelled axes in (\sref{fig:electron_density}{b and c}).
The resulting Abel inversions are not shown, as visually they appear very similar to \sref{fig:interferometry_results}{c and f}, but the phase maps (now in units of rad/cm) are used in the standard two-colour interferometry formulas to determine the electron and neutral density:\cite{Muraoka2000}
\begin{eqnarray}
n_e&=&\frac{\phi_1\lambda_1-\phi_2\lambda_2}{r_e(\lambda_2^2-\lambda_1^2)}\\
n_a&=&\frac{\phi_1/\lambda_1-\phi_2/\lambda_2}{\alpha_a(\lambda_1^{-2}-\lambda_2^{-2})}
\end{eqnarray}

Here the subscript 1 refers to the 532 nm interferometry and 2 refers to the 355 nm interferometry, $r_e=2.83\times10^{-13}$ cm is the classical electron radius and $\alpha_a=6.4\times10^{23}$ cm$^{-3}$ is the atomic polarisability for the dominant neutral species, carbon.\cite{Schwerdtfeger2014}
The electron and neutral densities calculated through these formula are shown in \sref{fig:electron_density}{b and c}.

The electron density of the plasma jet takes the form of a broad plume (\sref{fig:electron_density}{b}).
The colour scale has been chosen to emphasise the features further away from the gun nozzle, near where the Thomson scattering was performed, at \z{5}.
As expected, the electron density drops off rapidly both in the $\pm x$ direction and in the $z$ direction.

The neutral density (\sref{fig:electron_density}{c}) has a region of unphysical negative neutral density around $(x,z)=(0,3)$ mm.
The cause of this is unclear, but it is important to note that it is not related to the value of $\alpha_a$, the static polarisability, which only determines the neutral density up to a constant.
One possible explanation is the varying atomic mix in the plasma plume --- the coaxial cable is made from a copper conductor with a plastic insulator which contains traces of fluorine.
If the ratio of these elements changes throughout the plume due to different mobilities or ablation rates, then the neutral polarisability could also change across the plume.
In particular, if there are atomic absorption resonances close to either laser probing wavelength, these would strongly alter the polarisability.

This negative neutral density has also been inferred using time-resolved continuous wave triature interferometry on this plasma gun, and has been reproduced in many previous experiments [see, for example, Ref. \onlinecite{Macdonald2015}].
These results underline the care needed when using two-colour interferometry.

Importantly, the neutral density along \z{5} is positive, and so we believe that the two-colour interferometry correctly infers the electron density in this region, and this electron density also agrees with that inferred from Thomson scattering, which will be discussed in the next section.

\subsection{Thomson Scattering: Electron densities and temperatures}

\begin{figure*}[t]
	\Fig{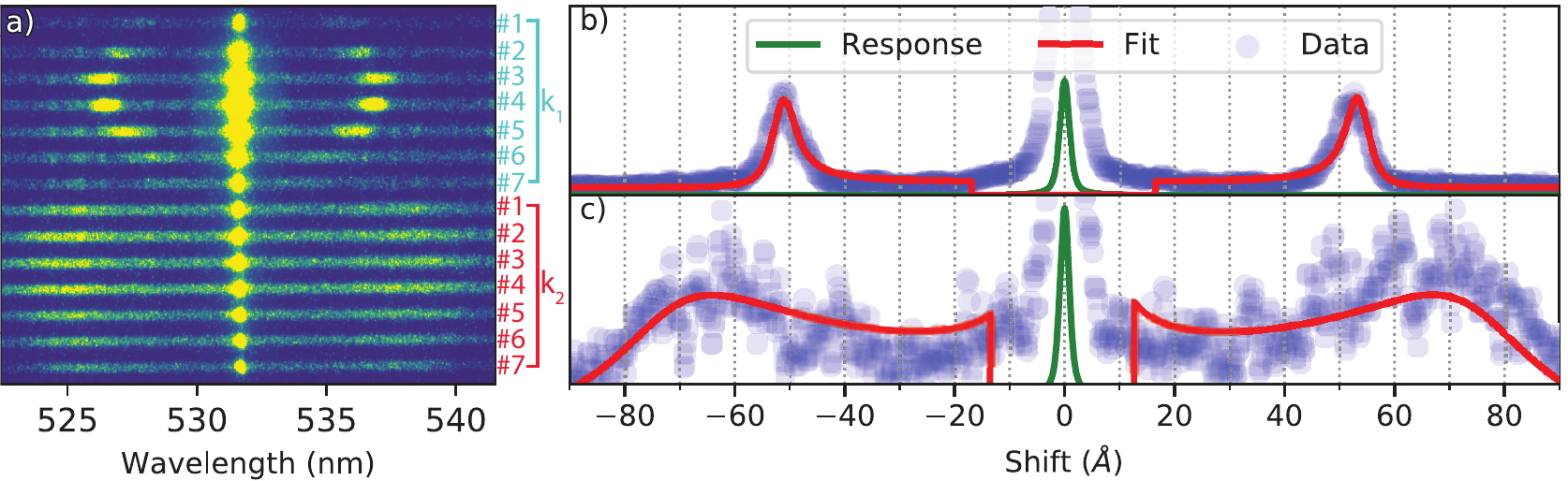}
	\centering
	\caption{Results from Thomson scattering at 940 ns after current a) Raw spectrogram showing variation of the scattering spectra in space. The top seven spectra correspond to the $k_{s1}$ scattering at \deg{24}, and the bottom correspond to $k_{s2}$ scattering at \deg{90}. Fibre 4 was aligned close to \x{0} for both fibre arrays. The change in electron density with position is clearly visible in both fibre arrays as the location of the electron plasma wave satellites are visible by eye. b) Fibre 4 from the $k_{s1}$ scattering direction, with the response of the spectrometer show in green, the ion feature (cut off) near 0 \AA{} shift and the electron plasma wave features at around $\pm50$\AA, showing scattering in the collective regime. A fit to the data is shown in red. c) Fibre 4 from the $k_{s2}$ scattering direction, response shown in green. Here the scattering is intermediate between the collective and non-collective regimes and the peaks are very broad.}
\end{figure*}

The Thomson scattering diagnostic collected scattered light from two angles in the ($x,y$) plane (\sref{fig:setup}{b}), simultaneously with the laser interferometry.
The scattered light was collected along \z{5} --- the locations of the fibres are shown in \sref{fig:electron_density}{b}.
The raw spectrogram is shown in \sref{fig:thomson_results}{a}, with the fourteen fibres visible as bright horizontal bands, and the unshifted laser light as series of bright regions down the centre of the image.
The top seven bands correspond to the seven fibres aligned in the $k_{s1}$ direction, at \deg{24} from the probing laser, and the bottom seven are from the fibres in the $k_{s2}$ direction at \deg{90}.
The spatial extents covered by the two fibre arrays are different due to the different angles --- the $k_{s1}$ array spans an extent $1/\sin{24^\circ}=2.5$ larger than the $k_{s2}$ array, and the collection volumes are elongated by the same factor in the $x$ direction.
This means that the locations of the scattering volumes do not match up for the two arrays, except for fibre 4, which was aligned close to $(x,z)=(0,5)$ mm for both fibre arrays.
In a separate shot we also collected the self-emission spectrum for the plasma with the same time after current start, exposure, camera gain etc.
The atomic lines and continuum emission were smaller by a factor of over five compared to the Thomson scattering signal shown in \sref{fig:thomson_results}{a}.

A great deal of information can be determined by simply looking at \sref{fig:thomson_results}{a}.
The most visually striking feature are the bright regions close to 532 nm, the initial laser frequency. 
This includes contributions from stray light (light scattered from metal surfaces), Rayleigh scattering from neutrals (much lower cross section than electrons, but can be significant for partially ionised plasmas) and the ion feature (light scattered from electrons which Debye shield the ions, and hence determined by the ion properties).
Scattering from the electron feature occurs at larger shifts from the initial wavelength, due to the higher mobility of free electrons.
It is not possible to simultaneously resolve the electron and ion feature on our spectrometer due to the inverse relationship between resolution and bandwidth, and so we focus on the electron feature.
This means we do not need to consider contributions from stray light and Rayleigh scattering, which only occur close to the initial wavelength.

The electron feature is very clear in the $k_1$ fibres, forming a pair of evenly split satellites in each fibre.
The shift of each electron plasma wave satellite is a function of both the electron density and electron temperature, as shown in eqn.  \ref{eqn:eqw_dispersion}.
The satellites are at the largest shift in fibres 3 and 4, close to the centre of the plasma jet, and become closer to the initial wavelength as the collection volumes move further from the centre of the jet.
This corresponds to the decrease in electron density seen in the interferograms in \sref{fig:electron_density}{b}.

The electron feature can also be seen in the $k_2$ fibres, which are at a larger angle, giving a smaller value of $\alpha=1/k\lambda_{De}\approx 1$, and hence are closer to the non-collective regime $\alpha<1$.
For true non-collective scattering, we would observe a Gaussian profile, as the scattering comes from individual electrons within the Maxwellian distribution without collective effects.
Instead, we observe satellites which are far broader, but occur at larger shifts than the $k_1$ satellites due to the more significant contribution of the plasma temperature to the dispersion relationship for the $k_2$ vector (as in eqn.  \ref{eqn:eqw_dispersion}.)
The trend in shift with fibre is less clear in these fibres for two reasons: firstly, the smaller spatial range covered by this fibre array and secondly, the significant contribution of the temperature (which barely changes over the range of volumes, see next subsection) to the overall Doppler shift of the satellites.

The overall spectrum for each fibre is found by integrating in the vertical direction over the 17 pixels closest to the centre of the emission. 
The spacing between fibres is 55 pixels, so this procedure avoids cross talk from other fibres contaminating the spectrum.
This unshifted light is obtained for a given slit width from a separate shot in which the spectrometer gain is set to maximum and the probe laser is used in the absence of the plasma --- in this case the spectrum comes from stray light scattered from the focal point halo by nearby metal surfaces.
The response function of the spectrometer is found from fitting the unshifted light with a Voigt profile (the convolution of a Gaussian profile with width $\sigma$ and a Lorentzian profile with width $\gamma$), resulting in $\sigma=0.6$ \AA{} and $\gamma=0.5$ \AA{}, which gives a FWHM of 2 \AA{}.
This response function is used to convolve the results from a multi-parameter fit in which specific parameters can be fixed or varied by the fitting routine.\cite{Hare2017a}
The data and model are multiplied by a ``notch filter'' to avoid fitting to the ion feature, which is not resolved in these experiments.

\sref{fig:thomson_results}{b} shows the spectrum for fibre 4 in the $k_{s1}$ array, shown with translucent blue circles.
The response function of the spectrometer is shown with a solid green line.
The red line shows the best fit, allowing the electron density, electron temperature and amplitude to vary --- the notch filter is clear here, and fitting does not occur within this region.
Note that the electron plasma wave satellites are much broader than the response function of the spectrometer, and so we are capable of resolving broadening due to electron thermal motion (temperature) or other effects.

The spectrum for the same scattering volume (fibre 4) from the $k_{s2}$ array is shown in \sref{fig:thomson_results}{c}.
The signal for this scattering angle has a lower signal to noise ratio due to the increased broadening.
For fitting purposes, we fixed the electron density for this fit to the value given from interferometry --- as the dispersion relationship is sensitive to both $n_e$ and $T_e$ for this scattering geometry, these two parameters are strongly inversely correlated in the fit and so a constraint must be added using prior knowledge from another diagnostic.
Hence for this fit, the electron temperature and the amplitude of the signal were the free parameters.

\begin{figure}[h]
	\Fig{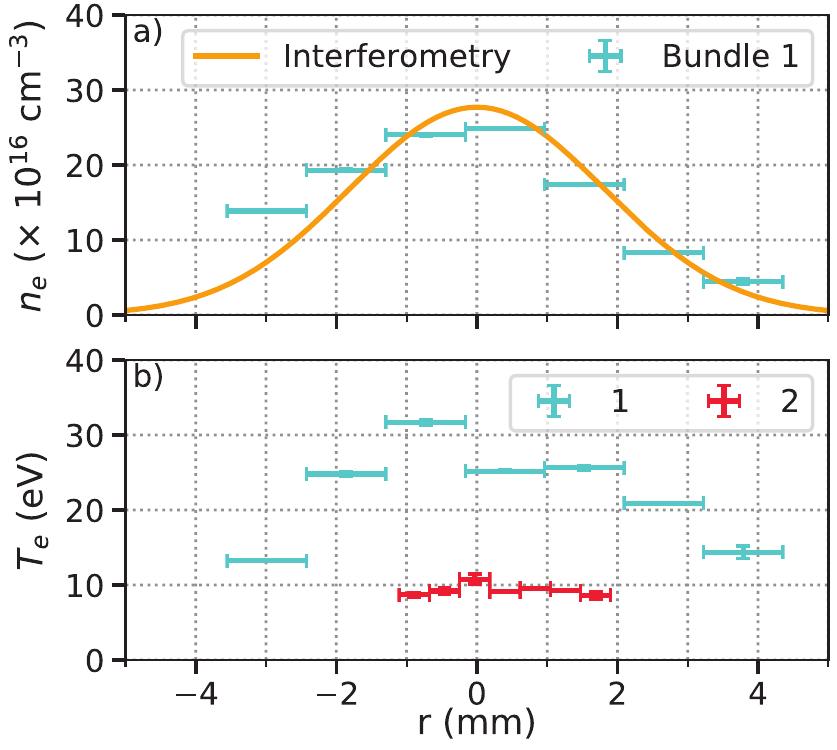}
	\centering
	\caption{Profiles of density and electron temperature from interferometry and Thomson scattering. a) Line-out of electron density along \z{5}, averaged over $\pm$ 0.2 mm perpendicular to the line-out (solid line, orange) and density inferred from collective Thomson scattering through shift of electron plasma waves in $k_1$ array (markers, cyan). The x error bars represent the horizontal extent of the collection volume. b) Electron temperatures inferred from broadening of electron plasma wave feature for $k_1$ (cyan) and $k_2$ (red). The inferred temperatures are significantly different. Note that the volumes from which light is collected by $k_1$ and $k_2$ are not the same due to the angles at which the light is collected.}
\end{figure}

The results of these fits for each of the seven fibres in the two fibre arrays are shown in \sref{fig:thomson_profiles}.
In \sref{fig:thomson_profiles}{a}, a line-out of
 the electron density taken along \z{5\pm0.5} is shown in orange.
Overlaid is the electron density obtained from fitting the fibres from the $k_{s1}$ array, allowing $n_e$ and $T_e$ to be varied.
The horizontal error bars here indicate the spatial extent of the collection volume.
The vertical error bars are provided by the fitting code, but provide an underestimate of the true error due to the inverse correlation between $n_e$ and $T_e$, which was usually around -0.8.
The electron density inferred from interferometry is closely matched by that inferred from the collective Thomson scattering of the electron feature.

\sref{fig:thomson_profiles}{b} shows the electron temperatures inferred from the two scattering directions.
The results from the fibres in the $k_{s2}$ direction are shown in red.
To fit these fibres, the electron density was set using the mean electron density within the collection volume, as determine by interferometry (\sref{fig:thomson_profiles}{a}).
The horizontal error bars again indicate the spatial extent of the collection volume, and the vertical error bars come from the fitting routine (with no correlation to $n_e$, which is fixed).
The electron temperature profile is flat, in the range 9--11 eV across the 3 mm spanned by this fibre array.

The results from the $k_{s1}$ array are in marked contrast to those from $k_{s2}$, and are shown in cyan.
The temperature inferred is significantly higher, reaching above 30 eV in the same spatial range in which the $k_{s2}$ array infers $\sim$10 eV.
This significant discrepancy is the focus of the remainder of this paper.

\section{Discussion}\label{sec:discussion}

The agreement between the electron densities from Thomson scattering and interferometry is unexpected but reassuring, especially in light of the issues of negative neutral density seen in some regions.
More surprising, however, is the difference between the temperatures inferred from two different angles, and by a factor of up to three.
Clearly there is some additional peak broadening mechanism not accounted for by our simple analysis.
In this section we will consider three distinct mechanisms: the finite aperture effect, density gradients, and collisional broadening.
By combining these effects we can obtain significant peak broadening for the electron plasma wave satellites.

\subsection{The Finite Aperture Effect}
\begin{figure}[h]
	\Fig{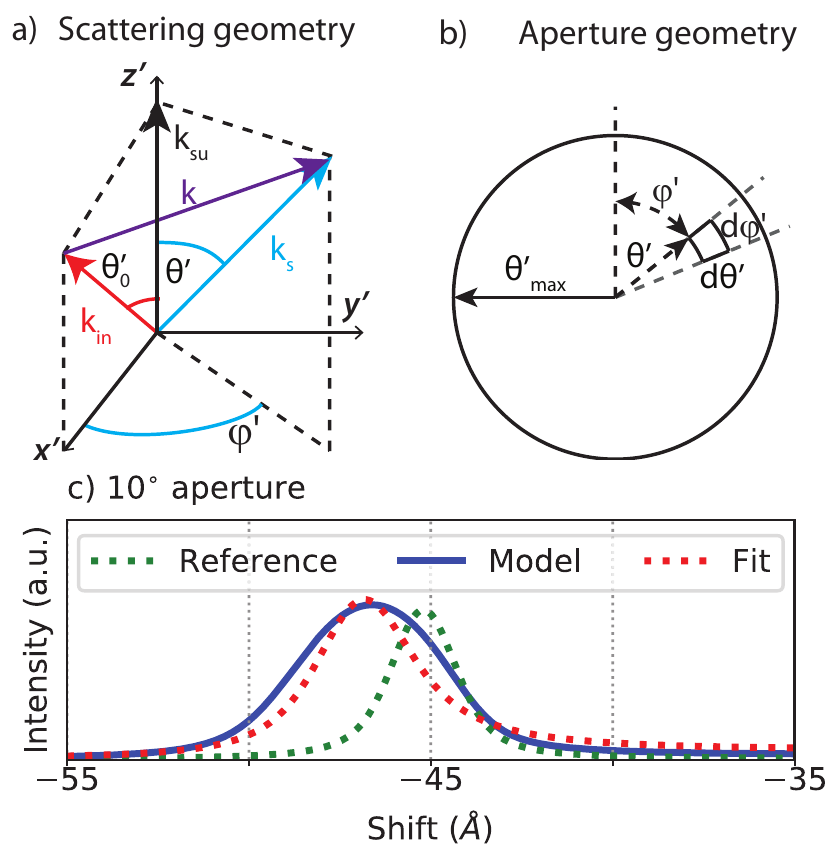}
	\centering
	\caption{
		a) The scattering geometry for finite aperture effects. Note that $(x',y',z')$ are not the same as the $(x,y,z)$ in \sref{fig:setup}{}. $k_{in}$ is the probing vector at a polar angle $\theta_0'$ in the $x'-z'$ plane, $k_{su}$ is the unshifted $k$-vector directed to the centre of the aperture, and $k_s$ is the scattering vector which enters the aperture at a polar angle $\theta '$ and an azimuthal angle $\phi '$. 
		b) The aperture geometry, with $\theta '$ and $\phi '$ corresponding to the scattering vector $k_s$ in a). The scattered spectrum is evaluated for $\theta '$=[0, $\theta_{max}'$] and $\phi '$=[0, $2\pi$]. 
		c) Simulated Thomson scattering spectra for a reference pinhole aperture with $T_e=9$ eV and $n_e=$ \xcmcubed{25}{16} (\textit{green dotted line}), the broadening due to a  
		a $\mathbf{f/3}$ aperture ($\theta_{max} ' =$\deg{10}) (\textit{blue solid line}), and a fit to this broadened spectrum which infers $n_e$=\xcmcubed{22}{16} and $T_e$=17 eV (\textit{red dotted line}).
	}
\end{figure}

The equations of Thomson scattering are derived for a single value of $k$, the resultant $k$-vector.
This implies that light is collected from only a single angle, which specifies an infinitesimally small aperture or a pinhole.
In any realistic Thomson scattering system, the apertures are finite --- this is necessary to collect more scattered light in order to improve the signal to noise ratio.
In our experiment, we placed a 50 mm diameter achromat at a distance of 150 mm from the plasma, so that this first optic subtends a cone with a range of $k$-vectors from \deg{14} to \deg{34} in the plane defined by the probing beam and the centre of the lens, and $\pm$\deg{10} out of this plane.
As the central scattering vector is only at \deg{24} and the scattering spectrum changes strongly as a function of $k$, this means the lens collects scattered light with very different spectra.

In order to calculate the size of this effect, we adopt the scattering geometry shown in \sref{fig:thomson_broadening}{a and b}.
This geometry is adapted from a more complex model intended to study the effects of finite apertures on the Thomson scattering spectrum for magnetised electrons.\cite{Hare2017a}
In \sref{fig:thomson_broadening}{a}, the scattering vector which points towards the centre of the lens is labelled $k_{su}$, and lies along the $z'$ axis (which is unrelated to the $z$ axis in our experimental configuration).
The probing laser $k_{in}$ lies in the $(z',x')$ plane at an angle $\theta_0'$ to $k_{su}$ --- this is the angle of the fibre array we have been quoting in the rest of this paper. 
The range of scattering vectors which can enter the lens are labelled $k_s$, and they are directed along a polar angle $\theta'$ with respect to the $z'$ axis and at an azimuthal angle $\phi'$ with respect to the $x'$ axis.

The aperture geometry is shown in \sref{fig:thomson_broadening}{b}.
The angles $\theta'$ and $\phi'$ are the same as those in \sref{fig:thomson_broadening}{a}, and $\theta_{max}'$ is the largest angle ray which can enter the lens --- \deg{10} for our system.
In order to calculate the total spectrum seen by the lens, the lens is broken up into small areas $dA=\sin{\theta'}d\theta' d\phi'$.
The spectrum is evaluated at each point on the aperture, and summed:
\begin{eqnarray}
S_{Tot}(k,\omega)=\int_0^{2\pi}\int_0^{\theta'} S(k(\theta', \phi'),\omega) \sin{\theta'}d\theta' d\phi'\label{eqn:finite_aperture}
\end{eqnarray}
For speed and clarity, the various $k$-vectors are treated as 3-vectors, and vector operations such as the dot product are used, rather than trigonometric identities.
For a typical calculation, we used eleven steps in $d\theta'$ and eleven in $d\phi'$.
The small area centred on $\theta'=0$ does not need to be calculated as $\sin \theta'=0$ here, which avoids calculating functions with singularities when magnetic fields are included.

After $S_{Tot}(k,\omega)$ is calculated, it is convolved with the response function of the spectrometer to take into account the intrinsic broadening from the instrument.
To demonstrate the effect of this broadening, we consider the parameters found for fibre 3 in the $k_{s1}$ array, in which the temperature inferred by $k_{s1}$ is 32 eV, much larger than the 9 eV inferred from the $k_{s2}$ array.
To begin with, we generate a synthetic spectrum for $n_e=$ \xcmcubed{25}{16} and $T_e$=9 eV, which is shown as a dashed green line in \sref{fig:thomson_broadening}{c}.
These parameters correspond to the average density with the collection volume of fibre 3 in the $k_{s1}$ array, and the average temperature inferred from the $k_{s2}$ array for this volume.
We note that for this collection angle the response function of the spectrometer has the same width as this convolved synthetic spectrum, and so we cannot resolve temperatures of less than 9 eV using this spectrometer and fibre geometry.

We calculate a synthetic spectrum for a finite aperture with a half-angle of \deg{10} ($f/3$) and show the spectrum of the blue shifted electron plasma wave satellite, at around -45 \AA{} to the initial wavelength (blue line, \sref{fig:thomson_broadening}{c}).
The satellite is significantly broader than the reference result (which uses an infinitesimal or pinhole aperture), and it is shifted further from the initial wavelength.
This is due to the different intensities of the electron feature as a function of $k$-vector --- for these parameters, the intensity of the scattering at larger $k$-vectors (larger angles) is stronger, and so when summed together, the spectrum has more of this character.

To test how large an effect this can have on our experimental results, we then fit the resulting broadened spectrum with our standard Thomson scattering model, which gives $n_e$=\xcmcubed{22}{16} and $T_e$=17 eV --- the inferred density has actually reduced, but this has been compensated for by a significant increase in the inferred temperature, which overall results in the peak shifting further from the initial wavelength (see eqn. \ref{eqn:eqw_dispersion}).
The significant increase in inferred temperature for the $k_{s1}$ array shows the importance of considering finite aperture effects when the collection angle is large.

\subsection{Density gradients}

\begin{figure}[h]
	\Fig{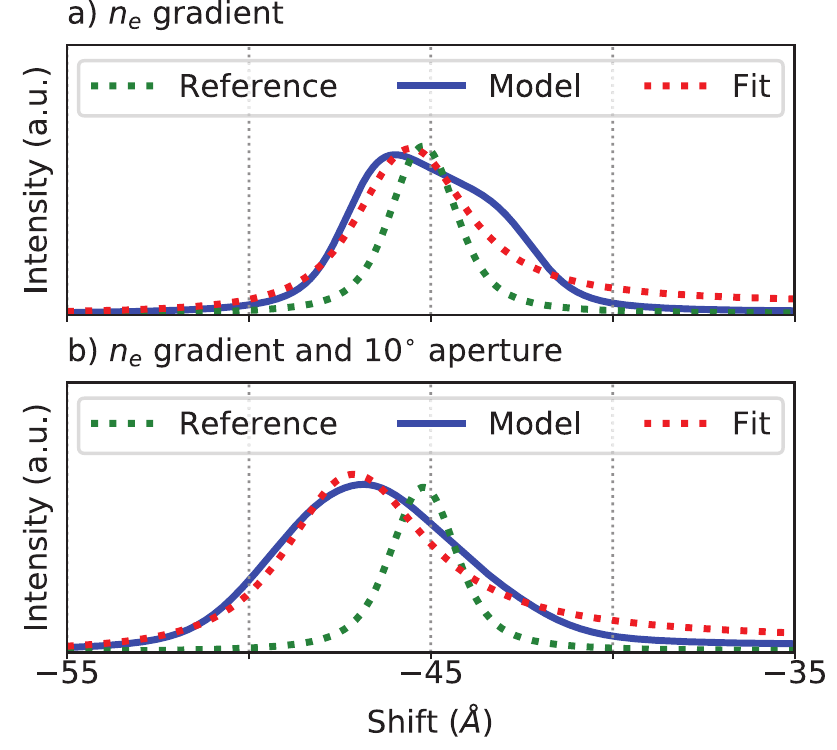}
	\centering
	\caption{
		a) Simulated Thomson scattering spectra for a reference pinhole aperture with $n_e=$ \xcmcubed{25}{16} and $T_e=9$ eV (\textit{green dotted line}),  
		a pinhole aperture with a range of $n_e$ taken from \sref{fig:thomson_profiles}{a} (\textit{blue solid line}), and a fit to this broadened spectrum (\textit{red dotted line}) which infers $n_e$=\xcmcubed{21}{16} and $T_e$=16 eV.
		b) The same reference spectrum (\textit{green dotted line}), a synthetic spectrum combining the finite aperture and density gradient effects(\textit{blue solid line}), and fit to the combined broadened spectrum (\textit{red dotted line}) which infers $n_e= $ \xcmcubed{21}{16} and $T_e=20$ eV.
	}
\end{figure}

A second important effect is the variation of plasma parameters throughout the collection volume, in the form of gradients in temperature and density.
As we have information on the nature of the density gradients from our interferometry diagnostic (\sref{fig:thomson_profiles}{a}), we will focus on this effect here, though temperature gradients may also play a role.
Again we consider fibre 3 in the $k_{s1}$ array, for which the interferometry shows densities between 21 and \xcmcubed{28}{16}.
Calculation of the broadening here is more straightforward --- we take the density at each point within the collection volume from the interferometry diagnostic, and calculate the spectral density function $S(k,\omega)$. 
Each spectra is calculated assuming $T_e=9$ eV, using the value from the $k_{s2}$ array.
The contributions are summed and then convolved with the response function of the spectrometer, as above.

The results of this calculation are shown in \sref{fig:thomson_gradient_combined}{a}.
The model (blue solid line) is very asymmetric, clearly showing the different intensities of the scattered light from different points within the collection volume and the best fit (red dotted line) gives $n_e$=\xcmcubed{21}{16} and $T_e$=16 eV.
Again, the reduction in density from the mean density within the volume is compensated by a significant increase in temperature.
The spectrum is significantly distorted from the usual form of the Electron Plasma Wave satellite (green dotted lined), and we do not observe this distortion in our results. 
However, the finite aperture effect and the density gradients work in concert, and so we must consider both together.

\subsection{Combined broadening from finite apertures and density gradients}

To combine these two effects, we calculate the finite aperture spectrum through eqn. \ref{eqn:finite_aperture} for each of the density points within the collection volume.
As before, the reference $T_e=9$ eV was used, and the densities were taken from \sref{fig:thomson_profiles}{a}.
This calculation required 7,500 spectra to be computed, each over 10,000 points in wavelength space.

The results are shown in \sref{fig:thomson_gradient_combined}{b}.
The model result (blue solid line) is significantly broader than the reference result (green dotted line).
The peak is also shifted further from the initial wavelength, and fitting (red dotted line) gives $n_e=$ \xcmcubed{21}{16} and $T_e=$ 20 eV, compared with 17 eV for the finite aperture and 16 eV for the density gradients alone.
Note that the spectrum is now quite symmetric around the electron plasma wavelength, and it is well fit by the model for a pinhole aperture at a single density.
It is difficult to distinguish experimentally between this broadened spectrum and the real, unbroadened spectrum given by a pinhole aperture in the absence of density gradients.

The inferred temperature has increased by more than a factor of two, demonstrating the importance of these two effects.
However, we still infer a temperature less than the 30 eV found by simply fitting the experimental data from the $k_{s1}$ array (\sref{fig:thomson_profiles}{b}).
This additional broadening could be due to small scale density gradients, not resolved by our interferometry, such as those caused by turbulence.
The effects of a finite aperture and density gradients were also calculated for the $k_{s2}$ array and found to be small --- these effects are far less important for this array as the both collection volume (density gradient effect) and fractional range of angles (finite aperture effect) are much smaller.

\subsection{Effect of collisions}

Collisions can also lead to broadening of the electron plasma wave features in Thomson scattering.
Using the model of Rozmus et. al (Ref. \onlinecite{Rozmus2017}), we calculate the broadening of the signal from a plasma at $T_e=9$ eV, $n_e=$ \xcmcubed{25}{16} and at an angle of \deg{24} --- the same conditions as the dashed green line in \sref{fig:thomson_broadening}{c}.
There is significant additional broadening of the unconvolved spectrum, but once the response function of our spectrometer is accounted for through convolution, the collisional broadening gives an effective $T_e=11$ eV.
This is less than the broadening from the finite aperture and density gradient effects, but does not depend on the collection optics, and so may be an important effect in other experiments.\cite{Davies2019}

\subsection{Asymmetry of the electron plasma waves}

\begin{figure}[t]
	\Fig{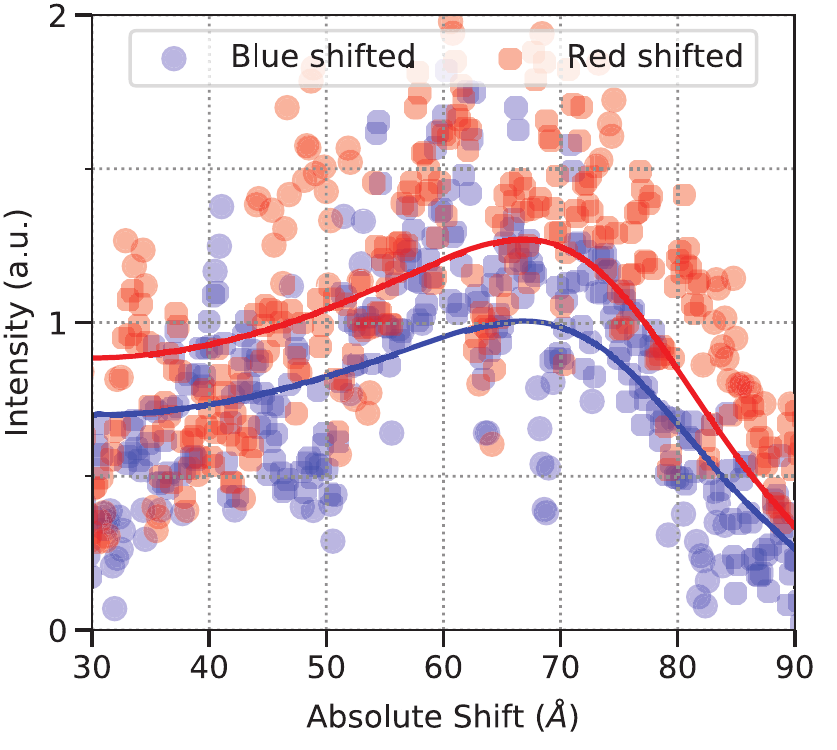}
	\centering
	\caption{Asymmetry of the blue and red shifted electron plasma waves. Data taken from fibre 4 of the $k_2$ array, with the red shifted spectrum reflected around the central wavelength so that it overlies the blue spectrum. Separately fitting symmetrised spectra gives a red shifted signal strength which is 27\% more intense than the blue shifted signal.}
\end{figure}

We note that there is a significant asymmetry in the intensities of the electron plasma waves in some collection volumes, as shown in \sref{fig:thomson_asymmetry}{} (and also visible in \sref{fig:thomson_results}{c}, the same spectrum.)
Here we have mirrored the blue shifted and the red shifted electron plasma waves around the initial wavelength and then fit them as two separate spectra to show the difference in peak intensity.
It is clear by eye that the red shifted wave is more intense than the blue shifted wave, despite the low signal to noise ratio, and the inferred intensity ratio is $I_{red}/I_{blue}\approx 1.3$.

Recent work on the Omega laser facility by Henchen et al. (Ref. \onlinecite{Henchen2018}) showed an asymmetry in electron plasma waves which was due to heat transport.
They observed measurable asymmetries when $\lambda_{ei}/L_T\sim10^{-2}$, where $\lambda_{ei}$ is the electron-ion mean free path and $L_T$ is the length scale of the temperature gradients.
In our experiments, we find $\lambda_{ei}/L_T\approx2\times10^{-3}$, which makes this plasma around an order of magnitude more collisional than that reported in Henchen et al.
Further work, including kinetic simulations, is therefore necessary to determine whether these kinetic heat transport effects could be important in the plasma gun outflow.

\section{Conclusions}

We have presented and compared temporally and spatially resolved measurements of the electron density and temperature in the outflow from a plasma gun.
Using two colour interferometry we determined the electron density in a region where the neutral density has a physical (positive) value, in contrast to other regions where the standard two-colour interferometry techniques give an unphysical (negative) neutral density.
These physical results suggest that additional sources of refractive index, such as resonances or ion polarisability, may be important in some regions of the plasma gun outflow.

We compared the results from the interferometry to those from Thomson scattering.
In the collective regime, the Thomson scattering spectrum contains resonances related to the electron plasma waves, with frequency shifts which are due to both the electron density and the electron temperature.
The electron densities measured from interferometry agreed with the measurements from Thomson scattering ($n_e\approx$ \xcmcubed{20}{17}), but the electron temperatures from Thomson scattering ($T_e\approx$ 20 to 30 eV) were much higher than previous measurements and simulations for plasma guns ($T_e\approx$ 5 to 10 eV).

We also obtained Thomson scattering spectra at a different angle, giving scattering in the non-collective regime.
The temperatures inferred from these spectra ($T_e\approx$ 9 eV) were in agreement with previous results, and in contrast to the other Thomson scattering measurements at a different scattering angle.
We attempted to reconcile these two measurements by taking a variety of broadening mechanisms into account, such as finite aperture effects, density gradients within the collection volume and collisional broadening.
Together, these mechanisms can give a inferred temperature of $T_e\approx 22$ eV, which is a significant increase of over 100\%, but is significantly less than the measured discrepancy of 200\%.
Clearly these mechanisms are important and may be significant in other Thomson scattering experiments, but the remaining broadening mechanisms are not yet understood.

Finally, we showed there was a significant asymmetry in the intensities of the electron plasma wave resonances which are red or blue shifted from the initial wavelength.
This effect has been previously reported as due to kinetic effects in electron heat transport, though the plasma in this current paper is more collisional than in the previous report.
However, the low cost and high repetition rate of a plasma gun makes it an ideal object for further investigation, which could include resolving the ion feature of the Thomson scattering or magnetising the plasma flow.

\section*{Acknowledgements}

This work was supported in part by AWE Aldermaston PLC, the Engineering and Physical Sciences Research Council (EPSRC) Grant No. EP/N013379/1, by the U.S. Department of Energy (DOE) Awards No. DE-F03-02NA00057 and No. DE-SC-0001063 and by a grant from DOE Fusion Energy Sciences user FWP100182.

\bibliography{library}

\appendix

\section{Magic2: An interpolation code for interferograms}\label{sec:magic}

\begin{figure}[h]
	\Fig{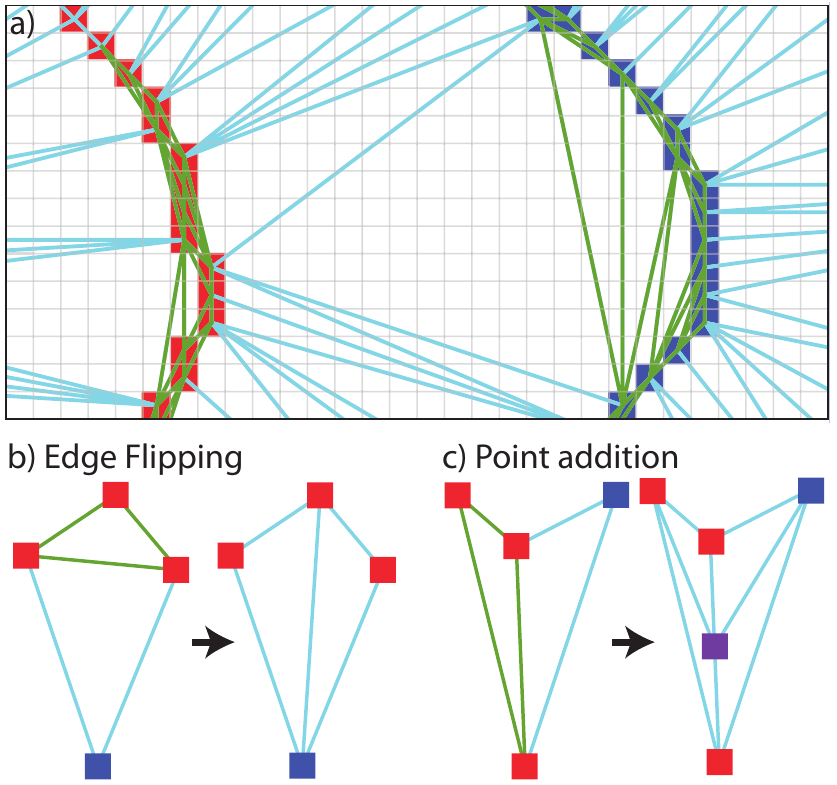}
	\centering
	\caption{Demonstration of the flat triangle problem in Delaunay triangulations, and some solutions. 
		a) Two contours (red and blue) on a pixel grid. Results of the Delaunay triangulation shown as lines, coloured blue for non-flat edges and green for flat edges.
		b) Flat triangle removal by edge flipping. A flat triangle (green edges) has all its vertices on the same contour (red points). It's neighbour has one vertex on a different contour, so is non-flat. By flipping the interior edge, we create two non-flat triangles.
	c) Flat triangle removal by vertex addition. For concave triangle pairs, it is not possible to edge flip, as flipping the interior edge results in an edge outside the two triangles. Instead, we add a new vertex along the interior edge, and interpolate its height from the existing vertices.}
\end{figure}

To process the interferograms presented in this paper, we use the newly written Magic2 code.
Magic2 is a reimplementation of MAGIC,\cite{Swadling2013} and in this appendix we will discuss the algorithms used to provide a faithful reconstruction of the phase map from the interferometry.
The code is freely available online at \newline \url{https://github.com/jdranczewski/Magic2}.

First, the fringes in the vacuum and shot interferograms are skeletonised into single pixel black lines on a white background.
This skeletonisation can be done manually, by tracing over the constructive and/or destructive interference fringes in software such as Adobe Photoshop, or it can be performed automatically with Fourier or Wavelet transform based techniques, or more advanced methods.
In practice, automated techniques require some manual correction due to artefacts such as self-emission, dust spots, Schlieren or shadowgraphy effects or loss of contrast due to laser beam profile.
Due to these effects, it is highly desirable that an experienced physicist interprets and checks the interferograms and phase maps, rather than relying on an entirely automated procedure, as this can prevent easily avoidable errors and artefacts.

The skeletonised interferograms are loaded into Magic2, which detects the fringes and presents them to the user for numbering via a GUI.
Using the numbered contours, a Delaunay triangulation is then performed, which divides up the interferogram into a set of triangles, which have vertices lying on the contour pixels.
At this point, it is possible to interpolate the Delaunay triangulation to quickly check whether the fringe numbering has been performed correctly.
However, there will be ``flat'' triangles present in the Delaunay triangulation, which have all three vertices with the same value --- see the green edged triangles in \sref{fig:magic2}{a}.
Interpolation within these flat triangles will lead to artificial plateaus, which are artefacts of the interpolation procedure and do not accurately represent the phase map.

Magic2 uses two methods to remove flat triangles: edge flipping and vertex addition.\cite{Ware1998a}
Triangles in Magic2 are objects which hold a list of pointers to their neighbouring triangles, and which of these neighbours are flat. 
This allows for iterative fixing of flat features without keeping a global list of flat triangles.
When Magic2 finds a flat triangle, it selects a non-flat neighbour --- if the pair forms a convex shape, it flips the inner edge to make two non-flat triangles (\sref{fig:magic2}{b}).
For concave triangle pairs, it is not possible to edge flip, as the new edge would lie outside the triangle pair.
Here we add a new vertex to the centre of the interior edge of the triangle pair. (\sref{fig:magic2}{c})
The value of this vertex is determined by a heuristic:

\begin{equation}
N = \frac{|B_1N|\cdot R + |R_2N|\cdot B}{|B_1N| + |R_2N|}
\end{equation}
where $|B_1N|$ is the distance between the new point $N$ and the blue vertex, $|R_2N|$ is the distance between the new point and either of the red vertices on the shared edge, and $B$, $R$ and $N$ are the value of the blue, red and new vertices respectively. 
This heuristic is used in the original version of MAGIC, but has not previously been published.

For each triangle, the barycentric coordinates are calculated for those pixels inside the rectangle which contains that triangle. For pixels outside the triangle, at least one of the coordinates has a negative value and so these pixels can be discarded, otherwise the calculated weights are used to interpolate a value. This method therefore finds the correct pixels for each triangle (a linear operation), which is much faster than finding the correct triangle for each pixel, as was done in the previous version of the code.

Once the vacuum and shot interferograms have been skeletonised, loaded, numbered, triangulated and interpolated, the resulting phase maps can be used to find the electron and neutral atom densities.
The interpolation code is of course general, and could be applied to interferograms generated using any technique (such as those in surface metrology), producing maps of phase which can then be interpreted appropriately for the given technique.

\end{document}